\documentclass[aps, twocolumn,noshowpacs,preprintnumbers,amsmath,amssymb]{revtex4}

\usepackage{verbatim}
\usepackage{wasysym}

\usepackage{graphicx}
\usepackage{dcolumn}
\usepackage{bm}
\usepackage{wrapfig}

\begin{document}

\title{Supramolecular interaction of singlewall carbon nanotubes with a functional TTF-based mediator probed by field-effect transistor devices}

\author{Annette Wurl}
\author{Sebastian Goossen}
\affiliation{Institute of Physical Chemistry, University of Hamburg, Hamburg, Germany}
\author{David Canevet}
\affiliation{Laboratoire MOLTECH-Anjou, UMR CNRS 6200, Universite d'Angers, 2 Bd Lavoisier, 49045 Angers Cedex, France}
\affiliation{IMDEA Nanociencia, Fco. Tomas y Valiente 7, 28049 Madrid, Spain}
\author{Marc Salle}
\affiliation{Laboratoire MOLTECH-Anjou, UMR CNRS 6200, Universite d'Angers, 2 Bd Lavoisier, 49045 Angers Cedex, France }
\author{Emilio M. Perez}
\affiliation{IMDEA Nanociencia, Fco. Tomas y Valiente 7, 28049 Madrid, Spain }
\author{Nazario Martin}
\affiliation{IMDEA Nanociencia, Fco. Tomas y Valiente 7, 28049 Madrid,  Spain }
\affiliation{Departamento de Quimica Organica, Facultad de Quimica, Universidad Complutense, 28040 Madrid, Spain}
\author{Christian Klinke}
\affiliation{Institute of Physical Chemistry, University of Hamburg, Hamburg, Germany }

\begin{abstract} 

The supramolecular interaction between individual singlewall carbon nanotubes and a functional organic material based on tetrathiafulvalene (TTF) is investigated by means of electric transport measurements in field effect transistor configuration as well as by NIR absorption spectroscopy. The results clearly point to a charge transfer interaction in which the adsorbed molecule serves as electron acceptor for the nanotubes through its pyrene units. Exposure to iodine vapors enhances this effect. The comparison with pristine carbon nanotube field effect transistor devices demonstrates the possibility to exploit charge transfer interactions taking place in supramolecular assemblies in which a mediator unit is used to transduce and enhance an external signal.

\end{abstract}

\maketitle

Carbon nanotubes (CNTs) are among the most promising candidates for new materials in the focus of nanoscience and nanotechnology. Possessing a wealth of unique chemical and physical properties, they show great potential for a wide range of applications and continue to be of increasing importance in both material and life sciences \cite{1,2,3,4}. Apart from mechanical robustness and chemical stability, the most prominent phenomena in CNTs are their superior electronic transport properties which arise from reduced scattering and therefore less power dissipation in the sp$^{2}$ hybridized carbon lattice \cite{5,6,7}. The most prominent example of devices which sought to exploit these properties is the carbon nanotube field effect transistor (CNTFET) \cite{1}. Intentionally built in the course of research in promising candidates for post-silicon logic device technology, their inherent potential to act as chemical sensors was recognized soon \cite{2,8}.

First, sensitivity towards the electronic environment is a result of the switching mechanism as well as the device geometry of CNTFET which is based on contact gating instead of channel gating like in conventional MOSFET. In other words, an electric field, caused for example by molecular dipoles of adsorbed species or charge density changes in the substrate, can stimulate the response of CNTFET-based sensor devices by shifiting the threshold voltage Vth \cite{7,9,10}. These effects are therefore termed electrostatic gating. Second, singlewall CNTs (SWCNTs) themselves can interact electronically through charge transfer processes with a broad range of analytes \cite{8,9,11,12,13,14,15}. This causes a shift in the CNTFET transfer characteristic also because the Fermi level of the CNT is altered, which in terms of FET operation refers to channel doping. Third, additional charges inside or in close proximity to the SWCNT can affect the transistor performance through scattering events \cite{16}. 

Finally, the structural properties of SWCNTs make them especially well-suited for non-covalent association with extended (bio-) organic molecular structures. The large surface-to-volume ratio of the sp$^{2}$ hybridized all-carbon lattice enables $\pi$-$\pi$ stacking, van der Waals forces and electrostatic interactions\cite{17,18,19,20}. In studies which sought to investigate these associations in detail, though, a more complex picture of mutual influences emerged. Apart from geometry dependencies, synergistic optoelectronic effects can dominate, especially in conjugated polymer-SWCNT assemblies and hybrid materials \cite{21,22}. These findings could be applied to the rational design of functional hybrid CNT-based materials. 

Apart from conjugated polymers with desired side chain functionalities, small aromatic molecules are among the most widely used moieties in functionalization agents since they are ideal "anchor groups" for further non-covalent decoration of SWCNTs \cite{22,23,24}. Among the latter, pyrene is known to have a very high affinity to the SWCNT surface \cite{25}.

Here, we present a model system for efficient chemical sensors. In our study, iodine sensing is achieved by a CNTFET non-covalently associated with 2,3-Bis[N-(1-pyrenylmethyl)aminocarbonylmethylsulfanyl]tetrathiaful-valene \textbf{1} (Figure 1A) \cite{26,27}. Compound \textbf{1} is capable of intermolecular van der Waals and sulfur-sulfur interactions as well as $\pi$-$\pi$ stacking and hydrogen bonding allowed by various chemical functions (highlighted in Figure 1A). The latter are responsible for \textbf{1} to form, in various solvents, three-dimensional networks constituted of supramolecular nanofibers, affording physical gels \cite{26}. Upon solvent evaporation, xerogels are obtained out of solution in a concentration dependent manner due to supramolecular polymerization. The organogelating as well as charge transport properties of xerogels derived from \textbf{1} in different solvents have been described, either as a neutral material or in the oxidized state \cite{27}. Incorporation of small amounts of SWCNTs (<0.1\% w) along the gelation process leads to xerogels presenting a significant increase in conductivity by 4-6 times. Considering this very low percentage, this result supports a structuring effect, which is promoted by templation of the supramolecular polymerization with the SWCNTs. This phenomenon leads to highly organized assemblies with significantly modified electronic properties. Now we investigate the opposite relation, namely the influence of \textbf{1} on the properties of SWCNTs at the single SWCNT level.

\begin{figure}[ht]
  \centering
  \includegraphics[width=0.4\textwidth]{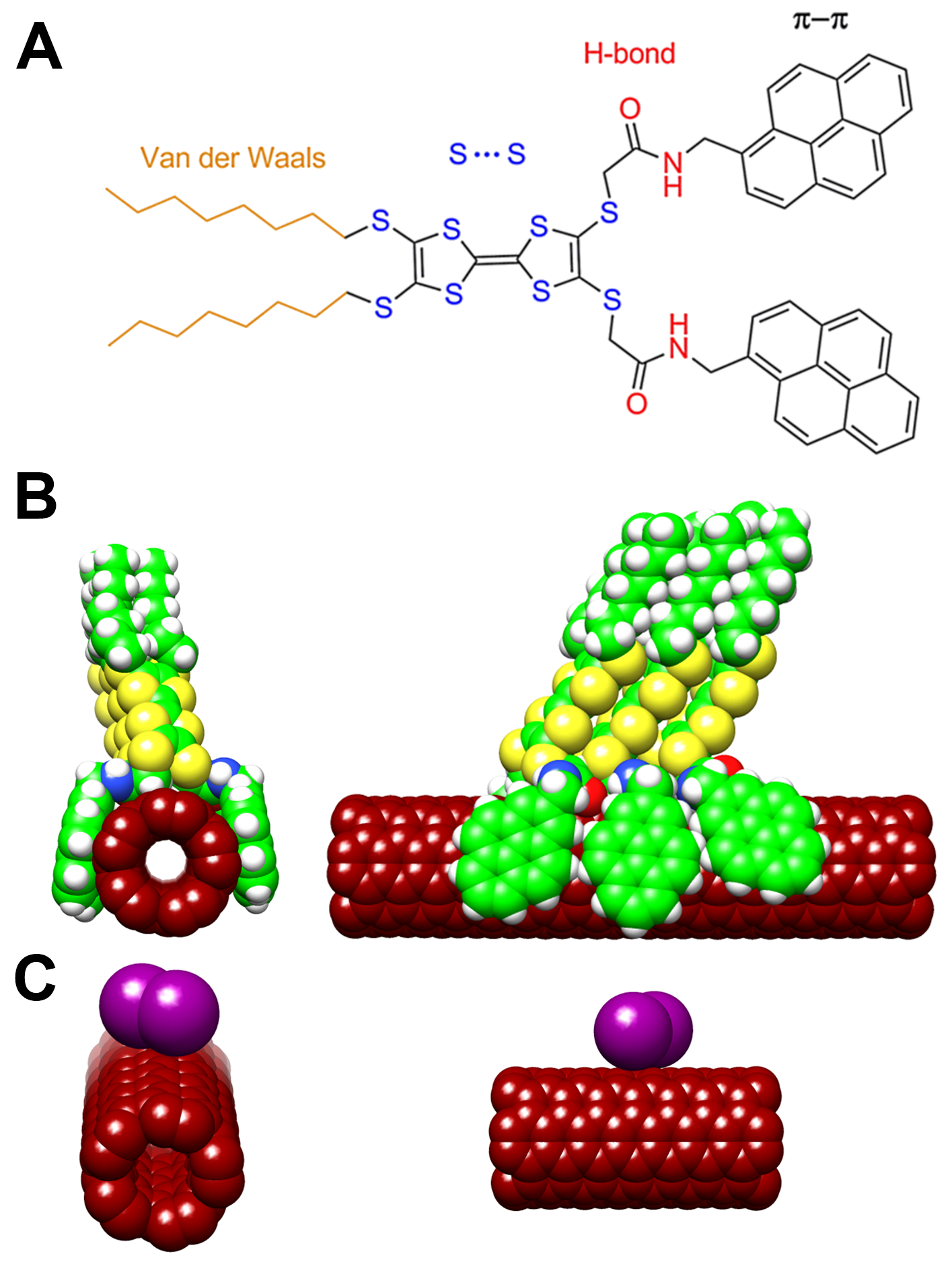}
  \caption{\textit{A) Structure of tetrathiafulvalene-based mediator \textbf{1} (functional groups highlighted in color), B) Front and side views of \textbf{1} assembled on a (5,5) nanotube, C) Iodine over a carbon nanotubes. B) + C) are geometry-optimized using the semi-empirical method PM3/vdW.}}
\end{figure}

The redox properties of tetrathiafulvalene (TTF) and its derivatives are well established, and have been exploited for long in the preparation of electroconducting salts or switches \cite{28,29}. In particular, a common feature of TTF-based molecular structures lies on two stable cationic oxidation states formed at rather low redox potentials that correspond to the radical mono-cation (TTF$^{+}$) and the di-cation (TTF$^{2+}$). The stability of these states partially stems from a gain in aromatization energy of the $\pi$ system of TTF \cite{28,30,31}. This gain is also responsible for the low oxidation potentials of most TTF derivatives and allows for their oxidation to the radical cation state with soft oxidizing agents, such as iodine.

Iodine is also known to act as a p-dopant for CNTs \cite{32,33,34}. Charge transfer interactions in mats of crystalline SWCNT ropes and thin films were investigated by Raman spectroscopy, X-ray diffraction, electrical transport data and UV-Vis absorption spectroscopy, respectively. In all of these studies, iodine was found to form charged linear chain complexes, namely (I$_{3}$)$^{-}$ and (I$_{5}$)$^{-}$. No neutral I2 could be detected, clearly indicating a transfer of electronic charge from the CNT lattice to iodine. In the case of SWCNT mats and multiwall CNTs (MWCNT), doping was carried out by immersing samples in molten iodine for prolonged periods of time. As for SWCNT-based thin films, they were incubated with iodine vapors at elevated temperature (320 K). In any case, electrical transport was investigated via dc resistance measurements. To our knowledge, no data exists for CNTFET doped by iodine in any form. The model system described here represents a new concept of effective chemical sensors in which a chemical reaction between the analyte, iodine, and a mediator system, multi-functional molecular material \textbf{1}, amplifies the response of a CNTFET through variations in its electronic environment \cite{14}.

The two pyrenyl units of \textbf{1} guarantee excellent attachment to the SWCNT surface via $\pi$-$\pi$ stacking. This point was confirmed by the clear modification of the SWCNT electronic properties upon supramolecular functionalization with \textbf{1}. Figure 2 shows the transfer characteristics of a CNTFET at VDS = -1 V and Vg in the range between -10 and +10 V of CNTFET before and after incubation with \textbf{1}, respectively. In the beginning all transistors show p-type behavior characteristics for CNTFET working at ambient conditions \cite{35}. In the presence of \textbf{1}, a shift of Vth of 5.0 V to more positive gate voltages as well as an increase in the ON state current by a factor of 1.54 is visible. The shift of the IDS-Vg characteristic shows that \textbf{1} acts as an electron acceptor upon association with SWCNTs, thereby altering the Fermi level towards the valence band edge, which results in a lowered barrier for hole conduction. The increase in lateral saturation current, on the other hand, can be attributed to the influence of \textbf{1} on the metal work function of the contacts and therefore the Schottky barrier height through dipole interactions \cite{9,36}. This explanation is further confirmed by the fact that the inverse sub-threshold slope, $S = dVg / d [ log(IDS) ]$ increases upon incubation of CNTFET devices with \textbf{1} from 1.85 V/dec to 2.83 V/dec (after oxidation with iodine: 4.51 V/dec). Due to the size of \textbf{1}, though, this must not be interpreted as diffusion into the contact region between the palladium electrodes and the SWCNT. \textbf{1} does also not act as a source of scattering for charge carriers inside the channel as this would lead to a decrease of the lateral current.

The hysteresis occurring in all measurement data is caused by mobile charges present in defect sites of the silicon dioxide substrate. While its width differs between different devices, the hysteresis is not affected significantly by any of the applied doping procedures when comparing data of the same device.

\begin{figure}[ht]
  \centering
  \includegraphics[width=0.4\textwidth]{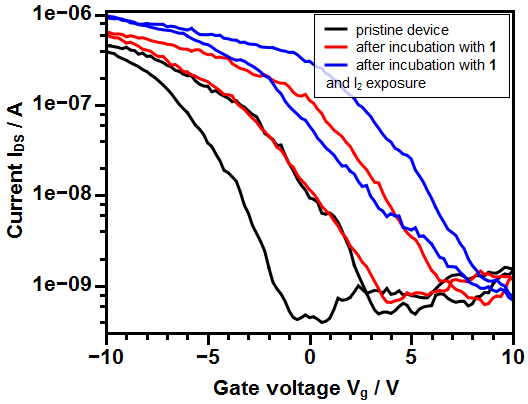}
  \caption{\textit{Typical CNTFET transfer characteristics at VDS = -1 V before (black) and after (red) incubation with molecular material \textbf{1} as well as after exposure to iodine vapor (blue).}}
\end{figure}

The effects of \textbf{1} on the transfer characteristics of CNTFET devices are interesting in terms of the functional moieties present in this material: Despite the TTF group, reputed as a donor unit, the overall molecular structure obviously is capable of serving as an electron acceptor for SWCNT. After iodine exposure, both effects on the transistor transfer characteristic described above are increased again: A stronger p-shift is evidence for the expected decrease in electron density throughout the delocalized $\pi$-system upon oxidation. The increase in (hole) conductivity may partly stem from the oxidized form of \textbf{1} itself, as has been reported for thin films of this material \cite{27}, but since the OFF current is not increased, the majority of charge transport must still take place in the CNTFET. 

We calculated the field-effect mobility $\mu_{FE}$ for holes from the data of the forward sweep in Figure 2, using a classical approach where the SWCNT is treated as a metallic cylinder \cite{37,38}.

\begin{equation}
\mu_{FE} = \frac{L^{2}}{V_{DS} C} \cdot \frac{dI_{DS}}{dV_{g}} 
\end{equation}

with $L$ being the device channel length $L$ = 250 nm and $C$ the capacitance of the channel with respect to the back gate. The latter is obtained by \cite{39}

\begin{equation}
C = \frac{2 \pi \epsilon_{avg} \epsilon_{0} L}{\ln (2 + 4 t_{ox} / d) } 
\end{equation}

Here, $\epsilon_{avg}$ = 2.45 is the average of the dielectric constant above - air $\epsilon$(Air) = 1.0 - and  - below $\epsilon$(SiO2) = 3.9 - the nanotubes, $t_{ox}$ is the oxide thickness  $t_{ox}$ = 100 nm and $d$ is the diameter of the SWCNT $d$ = 1.2 nm. $C$ then takes on a value of $C$ = 5.86 $\cdot$ 10$^{18}$ F.

We notice an increase in mobility after adsorption of \textbf{1} from about 9 cm$^{2}$/Vs to 11 cm$^{2}$/Vs. The oxidation further increases the mobility to about 13 cm$^{2}$/Vs.

In order to get a more detailed picture on the nature of interaction between \textbf{1} and SWCNT, we estimated the number of adsorbed molecules $N$ from the charge contribution, combining experimental and theoretical results. On the experimental side, the total charge $\Delta Q$ induced by \textbf{1} can be calculated from the shift in the threshold voltage: after adsorption of \textbf{1} $\Delta V_{th}$ = 5.0 V and after oxidation with iodine further $\Delta V_{th}$ = 6.0 V (the total shift is 11.0 V), between the respective measurements $\Delta Q = C \cdot V_{th}$. We obtain values of 2.93 $\cdot$ 10$^{17}$ C for the comparison of the cases before and after incubation with \textbf{1}, and 3.52 $\cdot$ 10$^{17}$ C for the comparison of the cases after incubation and after iodine exposure (the total charge transfer is 6.45 $\cdot$ 10$^{17}$ C). $\Delta Q$ then allows to estimate the increase in hole density per unit length $\Delta p$ inside the SWCNT that results from the charge transfer interaction with \textbf{1}:

\begin{equation}
\Delta p = \frac{\Delta Q}{e L} = \frac{C \Delta V _{th}}{e L}
\end{equation}

After incubation, an additional density of 0.732 nm$^{-1}$ is found and after oxidation by iodine a further increase of hole density $\Delta p$ of 0.878 nm$^{-1}$  is measured (the total hole density is 1.609 nm$^{-1}$ ). In comparison, density functional calculations show that 0.266 and 0.989 electrons are detracted from the nanotube by one molecule before and after oxidation, respectively. Thus, a reasonable agreement with the experiment is found. 

To obtain an approximate visualization of the supramolecular organization of \textbf{1} on the SWCNT, we carried out semi-empirical calculations (PM3/vdW). As a model system, we utilized a (5,5) SWCNT of about 4 nm in length and three molecules of \textbf{1}. The geometry-optimized structure shows \textbf{1} with the pyrene units embracing the nanotube in a pincer-like fashion, with the TTF units forming stacks with S-S distances between 4.3 and 5.6 \AA. Most of the amide groups are engaged in intermolecular hydrogen bonds, and the alkyl chains are establishing van der Waals interactions (Figure 1B). All these observations are in agreement with the expected behavior for \textbf{1}, and the data reported previously \cite{26}. These simulations on the assembly of \textbf{1} on the CNT surface show that supramolecular assembly is possible and mediated by $\pi$-$\pi$ stacking of the pyrene moieties. The pyrene-pyrene distance between two adjacent moieties adsorbed on the surface of the nanotube amounts to roughly 8 \AA. Together with the charge transfer considerations, we can state that the entire nanotube is covered by molecules of \textbf{1}.

\begin{figure}[ht]
  \centering
  \includegraphics[width=0.4\textwidth]{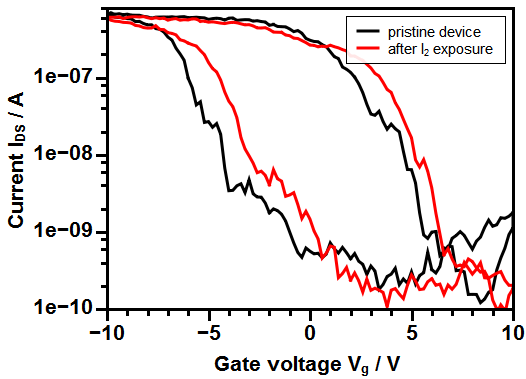}
  \caption{\textit{CNTFET Transfer characteristic at VDS = -1 V of bare devices before (black) and after (red) iodine exposure.}}
\end{figure}

We can clearly rule out the possibility that iodine itself is the cause of this charge-transfer phenomenon by data obtained from control devices where CNTFET were directly exposed to iodine vapors without the presence of \textbf{1}. A representative control experiment is shown in Figure 3. One can clearly distinguish between the effect of iodine on functionalized and non-functionalized devices: In the latter case, a weak p-shift and no increase in conductivity are observed. Also the simulations show a charge transfer of only 0.074 electrons per molecule. 

The results of the simulations show also that the adsorption of \textbf{1} and iodine on the nanotubes is of electrostatic nature. The minimal distance between the nanotube and \textbf{1} is comparatively large with 2.99 and 4.49 \AA. 

The results of Vis-NIR-absorption spectroscopy of SWCNT in solution support the explanation of the transport measurement data with charge transfer. A strong decrease of the S11 feature occurs in the spectrum of the sample incubated with \textbf{1}, an effect which is known in the literature for derivatives of pyrene \cite{25}.

In conclusion, we investigated the effect of supramolecular material issued from the association between \textbf{1} and a SWCNT. The influence of \textbf{1} on the electronic behavior of the nanotube was measured on the single SWCNT level in a CNTFET device configuration before and after association by recording transfer and output characteristics. A shift of the threshold voltage to more positive values in devices functionalized with \textbf{1} indicates transfer of electron density from the SWCNT (resulting in a shift of the Fermi level towards the valence band edge) to the organic compound. NIR absorption spectra of SWCNT associated with \textbf{1} in solution support this finding. An increase in conductivity in functionalized CNTFET, on the other hand, may be the result of an influence of \textbf{1} on the band alignment at the SWCNT-metal interface through dipole interactions. Upon exposure of functionalized devices to iodine vapors, both effects on the transfer characteristics become even more pronounced. Data obtained from control experiments in which non-functionalized CNTFET were exposed to iodine showed a similar trend, but an overall much weaker response. The work presented here serves as a model system for chemical sensors in which signal amplification occurs through a mediator system that interacts with a SWCNT electronically.

\section*{ACKNOWLEDGMENT}

Financial support from the Free and Hanseatic City of Hamburg in the context of the "Landesexzellenzinitiative Hamburg: Spintronics" is gratefully acknowledged. Financial support from MINECO (CTQ2011-25714) is gratefully acknowledged. EMP is also thankful for a Ramón y Cajal Fellowship, co-financed by European Social Funds.


\end{document}